\documentstyle[11pt,epsf]{article}
\newcommand{\beq}{\begin{equation}}
\newcommand{\eeq}{\end{equation}}
\newcommand{\beqa}{\begin{eqnarray}}
\newcommand{\eeqa}{\end{eqnarray}}
\newcommand{\beqan}{\begin{eqnarray*}}
\newcommand{\eeqan}{\end{eqnarray*}}

\newcommand{\ben}{\begin{enumerate}}
\newcommand{\een}{\end{enumerate}}
\newcommand{\bfl}{\begin{flushleft}}
\newcommand{\efl}{\end{flushleft}}
\newcommand{\ba}{\begin{array}}
\newcommand{\ea}{\end{array}}
\newcommand{\btab}{\begin{tabular}}
\newcommand{\etab}{\end{tabular}}
\newcommand{\bit}{\begin{itemize}}
\newcommand{\eit}{\end{itemize}}

\newcommand{\hs}{\hspace}

\def \g5 {\gamma_{5}}
\def \I {\hbox{ i}}
\def \W {\hbox{w}}
\def \D {\hbox{d}}
 \def\psl{\not{\hbox{\kern-2.3pt $p$}}} 
\def\Psl{\not{\hbox{\kern-2.3pt $P$}}} 
\def\ksl{\not{\hbox{\kern-2.3pt $k$}}}

\newcommand{\prepr}[1] {\begin{flushright} {\bf #1} \end{flushright} \vskip
1.5cm}
\newcommand{\titul}[1] {\begin{center}{\Large {\bf #1 } } \end{center} \vskip 1.cm}

\newcounter{muni}

\begin{document}
\vspace{.1cm}
\hbadness=10000
\pagenumbering{arabic}
\begin{titlepage}
\prepr{PAR-LPTHE-98-47\\ October 1998 }
\titul{Lepton flavor changing in neutrinoless $\tau$ decays.}
\vspace{5mm}
\begin{center}

{\bf Xuan-Yem Pham \footnote{\rm Postal address: LPTHE, 
Tour 16, $1^{er}$ Etage, 
4 Place Jussieu, F-75252 Paris CEDEX 05, France. \\
. \hspace{5mm} Electronic address : pham@lpthe.jussieu.fr }
}
\end{center} 
\vspace{0.5mm}
\begin{center}
{\large \bf \it
Laboratoire de Physique Th\'eorique et Hautes Energies, Paris \\
CNRS, Universit\'e P. et M. Curie, Universit\'e D. Diderot} 
\end{center}

\thispagestyle{empty}
\vspace{0.5mm}
\hspace{0.01cm} \large{} {\bf  \hspace{0.1mm} Abstract-}    
\normalsize
Neutrino oscillations, as recently reported by the Super-Kamiokande collaboration, imply that lepton numbers could be violated, and  $\tau^{\pm} \to \mu^{\pm} +\ell^{+} 
+\ell^{-}, \tau^{\pm} \to \mu^{\pm} +\rho^0$ are some typical examples. We point out that in these neutrinoless modes, the GIM cancelation is much milder with only a logarithmic behavior $\log (m_j /m_k)$ where $m_{j, k}$ are the neutrino masses. This is in sharp contrast with the vanishingly small amplitude $\tau^{\pm}\to \mu^{\pm} 
+ \gamma$ strongly
suppressed by the quadratic power  $(m_j^2-m_k^2)/ M_{\rm W}^2$. In comparison with the hopelessly small branching ratio B$(\tau^{\pm}\to \mu^{\pm} 
+ \gamma) \approx 10^{-40}$, the 
B$(\tau^{\pm} \to \mu^{\pm} + \ell^{+} 
+\ell^{-})$ could be larger than $10^{-14}$. The latter mode, if measurable, could give one more constraint to the lepton mixing angle $\sin 2\theta_{jk}$ and the neutrino mass ratio $m_j/m_k$, and therefore is complementary to neutrino oscillation experiments.

\vspace{10mm}
{\bf PACS numbers : 12.15.Ff, 12.15.Lk, 13.35.Dx, 14.60.Pq }
\end{titlepage}

\newpage

Evidence for the transmutation between the two neutrino species $\nu_\mu \leftrightarrow \nu_\tau$ is recently reported  by the Super-Kamiokande collaboration$^{(1)}$. As a consequence, neutrinos could have nondegenerate tiny masses and lepton numbers would no longer be conserved. Hence, besides the well known neutrino oscillation phenomena, neutrinoless $\tau$ decays, such as $\tau^{\pm} \to \mu^{\pm} + \gamma$, $\tau^{\pm} \to \mu^{\pm} +\ell^{+} 
+\ell^{-}$ and $\tau^{\pm} \to \mu^{\pm} +\rho^0$  could occur. 

The interest of the $\tau^{\pm} \to \mu^{\pm} +\ell^{+} 
+\ell^{-}$ and/or $\tau^{\pm} \to \mu^{\pm} +\rho^0$ modes is twofold. First, contrarily to the radiative case  $\tau^{\pm} \to \mu^{\pm} +\gamma$ which is damped by a vanishingly quadratic power$^2$, the typical lepton flavor changing 
amplitude $\tau^{\pm} \to \mu^{\pm} +\ell^{+} 
+\ell^{-}$ (or $\tau^{\pm} \to \mu^{\pm} +\rho^0$) is only suppressed by a smooth logarithmic term. Second, these $\tau$ decay modes, if measurable, are  ${\it complementary}$ to the neutrino oscillation experiments. They could give -- besides the lepton mixing angle $\theta_{ij}$ --  the ratio $m^2_j/m^2_k$, whereas neutrino oscillations  give the difference 
$|m_j^2 - m_k^2|$. By combining them, the absolute value of $m_j$ could be determined in principle.

Similarly to the CKM flavor mixing in the quark sector, let us assume that the neutrino gauge-interaction eigenstates $\nu_{\rm e}$, $\nu_\mu$ and $\nu_\tau$ are linear combinations of the three neutrino mass eigenstates $\nu_1$, $\nu_2$ and $\nu_3$ of nonzero and nondegenerate masses $m_1$, $m_2$ and $m_3$ respectively. Thus
\begin{equation}
\pmatrix{ \nu_{\rm e}\cr \nu_\mu\cr \nu_\tau\cr}
=\pmatrix { U_{{\rm e}1} & U_{{\rm e}2} & U_{{\rm e}3}\cr
                 U_{\mu 1} & U_{\mu 2 } & U_{\mu 3}\cr
                 U_{\tau 1}  & U_{\tau 2} & U_{\tau 3}\cr}
               \pmatrix {\nu_1\cr \nu_2 \cr \nu_3\cr} 
\equiv \; {\cal U}_{\rm lep}\;\pmatrix {\nu_1\cr \nu_2 \cr \nu_3\cr } , 
\label{eq:1} 
\end{equation}
where the $3\times 3$ matrix $\;{\cal U}_{\rm lep}\;$ is unitarity. Neutrino oscillation measurements give constraints usually plotted in the 
$(\sin 2\theta$,  $\Delta m^2= |m^2_i-m_j^2|)$ plane, where $\theta$ is one of the three Euler angles in the rotation matrix $\;{\cal U}_{\rm lep}\;$. 

The weak interaction effective Lagrangian for charged current of leptons can be written as
$${\cal L}_{\rm eff} ={{\rm G}_{\rm F} \over \sqrt{2}} L^{\dagger}_\lambda L_\lambda \;,$$ 
where the charged current $L_\lambda$ is
$$L_\lambda =\sum_{j =1}^{3} \overline{\ell}\gamma_{\lambda}(1-\gamma_5) \nu_j U_{\ell j}\;.$$
Here $\ell$ stands for e$^-$, $\mu^-$, $\tau^-$ and $\nu_j$ (with $j=1,2,3$) are the three neutrino mass eigenstates. For any fixed $\ell$, one has $\sum_{j}\vert U_{\ell j}\vert^2 =1$. For instance the $\nu_\mu$, operationally defined to be the invisible particle missing in the $\pi^+ \to \mu^{+} + \nu_{\mu}$, is initially a superposition of $\nu_1$, $\nu_2$ and $\nu_3$, in the same way as the K$^0$ meson produced by strong interaction, say by $\pi^- +{\rm p} \to {\rm K}^0 +\Lambda$, is initially a superposition of the mass eigenstates K$^0_{\rm L}$ and K$^0_{\rm S}$ with masses $m_{\rm L} \not= m_{\rm S}$. The nondegenerate masses are the origin of the oscillation phenomena of both neutrinos and neutral K mesons.

 In the most general renormalizable R$_{\xi}$ gauge, at one loop level to order $g^4$ -- where $g=e/\sin \theta_{\rm W}$ is the weak interaction coupling constant -- there are in all eighteen Feynman diagrams contributing to the neutrinoless decays $\tau^{\pm} \to \mu^{\pm} +\ell^+ +\ell ^-$ or $\tau^{\pm} \to \mu^{\pm} +\rho^0$; these modes are mediated by the Z and the photon, ten diagrams for the virtual Z, and eight for the virtual $\gamma$. Three of them are depicted in Figs.1--3.
The fifteen others, not shown here, are similar to Figs.1--3 in which 
the internal W$^{\pm}$ in loops are replaced in all possible ways by the "would be" Goldstone
 bosons $\Phi^\pm$, those absorbed by the gauge bosons W$^{\pm}$  to render them massive by the Higgs mechanism. The intermediate virtual photon is absent in Fig.1 and in Fig.1bis which is the analogue (not depicted here) of Fig.1 with W$^\pm$ replaced by $\Phi^\pm$. The  contributions of the mediated  neutral Higgs boson H$^0$ are negligibly small for both reasons (its mass and its couplings with the lepton $\ell$ pair or the up down quarks of the $\rho^0$) and can be discarded.

A careful examination of these eighteen diagrams shows that only Fig.1 and Fig.1bis provide the logarithmic behavior $\log (m^2_j/M^2_{\rm W})$, while the contributions of all other 16 diagrams are power suppressed as $(m^2_j/M^2_{\rm W})$, $(m^2_j/M^2_{\rm W})\times \log (m^2_j/M^2_{\rm W})$ and therefore vanishingly small.  The principal reason for the appearance of  the logarithmic $\log (m^2_j/M^2_{\rm W})$ term is that we are dealing in Fig.1 and Fig.1bis with two propagators of nearly massless fermions for which if the momentum transfer $q^2$ is much smaller than  $M_{\rm W}^2$ and consequently neglected, infrared divergences appear when mass of the {\it internal} fermion goes to zero$^{3, 4}$. We emphasize that Fig.1 and Fig.1bis are the only ones that contain an infrared divergence $\log (m^2_j/M^2_{\rm W})$, this fact has been noticed longtime ago in different contexts, for instance in the computation of the slope of the neutrino electromagnetic form factor$^{3}$, and the s-d-$\gamma$ induced coupling$^{4}$. Note however that compared to Fig.1, the contribution of Fig.1bis is damped by an additional 
$Mm/ M^2_{\rm W}$  factor because of the $\Phi$-fermion couplings, where $M$ and $m$ are respectively the $\tau$ lepton and muon masses. So actually only Fig.1 dominates. 

Due to the unitarity of the ${\cal U}_{\rm lep}$ reflecting the GIM cancelation mechanism, the  divergence as well as the $m_j$-independent finite part of the loop integral do not contribute to the decay amplitude because they are multiplied by $\sum_j (U_{\mu j}^* U_{\tau j}) =0$ when we sum over all the three neutrino contributions. Only the $m_j$-dependent finite part of the loop integral is relevant. This point is crucial, implying that we cannot neglect $m_j$ no matter how small $m_j$ is, otherwise we would get identically zero result after the summation over the neutrino species $j$. 

Our first task is to show that Fig.1 and Fig. 1bis actually give rises to the logarithmic $\log(m^2_j/M^2_{\rm W})$ term. This term could be equally guessed by approximating the W propagator with $\;$ i$/ M_{\rm W}^2$, the W mass plays the role of the loop integral momentum cutoff. Hence Fig.1 looks like the familiar  fermionic loop of the gauge boson self energy, or vacuum polarization. When $q^2 =0$ ($q$ being the four-momentum of the external gauge boson), the 
standard  
$\log (m_j^2/\mu^2)$ appears$^5$.
The following calculation of the diagram of Fig.1 confirms this expectation.

Let us write the one-loop effective $\tau$--$\mu$--Z transition of Fig.1 as $\overline{u}(p) \Gamma^{\lambda}_j(q^2) u(P)\varepsilon_\lambda(q)$, where $P$,  $p$ and $q=P-p$ are respectively the four-momentum of the $\tau$ lepton, muon and  virtual boson Z. Thus
\begin{equation}
\Gamma^\lambda_j(q^2) =\left({-\I g\over 4 \cos\theta_{\W}}\right)\left({-\I g\over 2\sqrt{2}}\right)^2 U_{\tau j} U_{\mu j}^* \int {\D^4 k\over (2\pi)^4} Y_j^\lambda (k, q)\;,
\label{eq:2}
\end{equation}
where
$$ Y_j^\lambda (k, q) = {\gamma^\rho (1-\gamma_5) \left[\I(\psl +\ksl +m_j)\right] \gamma^\lambda(1-\gamma_5)\left[\I(\Psl +\ksl +m_j)\right]\gamma^{\sigma} (1-\gamma_5) (-\I g_{\rho \sigma})\over \left[(k+p)^2-m_j^2\right]\left[(k+P)^2-m_j^2\right]\left[k^2-M_W^2\right]}\;, $$%
\begin{equation}
\hskip-2.2cm= 4 \I \;{\gamma^\rho \left[\psl +\ksl\right] \gamma^\lambda\left[\Psl +\ksl \right]\gamma_{\rho} (1-\gamma_5) \over \left[(k+p)^2-m_j^2\right]\left[(k+P)^2-m_j^2\right]\left[k^2-M_W^2\right]}\;.
\label{eq:3}
\end{equation}
The $\xi$ dependence in the W propagator of Fig.1 is canceled by the $\xi$ dependence of Fig.1bis (where $\Phi$ replaces W), so for simplicity, we take the $\xi=1$ Feynman--'t Hooft gauge at the outset.

Inserting $\Gamma^\lambda _j(q^2)$ inside $\overline{u}(p)$ and $u(P)$, making use of Dirac equations for these spinors and adopting  the standard Feynman paramerization for the denominator in $Y_j^\lambda (k,q)$, we get after the $ k$ integration
\begin{equation}
\Gamma^\lambda _j(q^2) ={\I g^3 U_{\tau j} U_{\mu j}^* \over 64 \pi^2 \cos\theta_{\W}}
\int_0^1 \D x \int_0^{1-x} \D y \;{{\cal N}^\lambda (q^2)\over {\cal D}_j(q^2) }\;,
\end{equation}
where
$$
{\cal N}^\lambda(q^2) = a \gamma^\lambda(1-\gamma_5) + b \gamma^\lambda (1+\gamma_5)
+c {(P+p)^\lambda\over M}(1+\gamma_5) + d {(P+p)^\lambda \over M}(1-\gamma_5) 
+e {q^\lambda\over M}(1+\gamma_5) + f {q^\lambda\over M}(1-\gamma_5)\,, 
$$%
\begin{equation}
{\cal D}_j(q^2)  = x\left[ M_{\rm W}^2 -M^2(1-x-y) -m^2y\right] -q^2 y(1-x-y) + m_j^2 (1-x) \;.
\label{eq:d}
\end{equation}
The form factors $a,\cdots, f$ are
$$
a= {\cal D}_j \log\left({{\cal D}_j \over M_{\rm W}^2}\right)  + M^2
\,x(x+y) + m^2\, x(1-y) - q^2 \,(x+y)(1-y)\;,\;b= M m\,x \;,\;c = - M^2\,x(x+y) $$%
\begin{equation}
d= M m \,x(1+y)
\;,\;e= - M^2\,(x+y) (x+2y-2)\;,\; f= M m \left[x(1+y) -2y(1-y)\right]\,.
\end{equation}
The $\log ({\cal D}_j /M_{\rm W}^2)$ in (6) is the finite part extracted from ${\cal D}_j^{-\varepsilon} \Gamma(\varepsilon) \approx [1 -\varepsilon \log ({\cal D}_j /M_{\rm W}^2)]\Gamma(\varepsilon) $ where $\varepsilon=2-(n/2)$ is the $n$-dimensional regularization parameter used to handle the ultraviolet divergence of (2).
As discussed, the $j$-dependence in ${\cal D}_j(q^2)$ as given by (5) is crucial to get nonzero result. On the other hand, since the $j$-independent terms $q^2 y(1-x-y) \leq q^2/4 \leq (M-m)^2 /4 \ll M_{\rm W}^2$ is much smaller than $M_{\rm W}^2$, it is useful to make an expansion of $\Gamma^\lambda _j(q^2)$ in power of $\eta\equiv q^2/M_{\rm W}^2$ by writing ${\cal D}_j(q^2)= {\cal D}_j(0) + {\cal O}(\eta)$, where  
\begin{equation}
{\cal D}_j(0) =  x\, [M_{\rm W}^2-m_j^2 -M^2(1-x-y) -m^2 y]+ m_j^2 .
\end{equation} 
This $\eta$ expansion simplifies the $x, y$ integrations and shows us that the relevent $j$-dependent part of $\Gamma^\lambda _j(q^2)$ has the following general form (the Lorentz index $\lambda$ is omitted for simplicity) :
\begin{equation}
\Gamma_j(q^2) = (A\eta +B) F(\eta, m_j) + G(m_j)\,, 
\end{equation}
 where $A, B$ are constant, and most importantly we note that $F(0, m_j) =\log \delta_j$, with $\delta_j \equiv m_j^2/M_{\rm W}^2$. The second term $G(m_j)$ only contains  $\delta_j$, $\delta_j\log\delta_j$ and $\delta_j^2\log\delta_j$ terms which are negligible for $m_j\to 0$.

The general structure in (8) emerges after the $x, y$ integrations of (4), using (7). This dominant $\log\delta_j$ manifests itself from the lower limit $x=0$ of the  $\int \int \D x \D y\;[1 
/{\cal D}_j(0)] $ integration. Note that  $\int \int \D x \D y\;[1 
/{\cal D}_j(0)] $ gives $\log\delta_j$, while $\int \int \D x\D y 
\;[x^n/{\cal D}_j(0)]$  gives  $\delta^n_j \,\log\delta_j$. Therefore, if there  exists terms ${\it  independent}$ of $x$ in the numerator ${\cal N}^\lambda(q^2)$, then these $x$-independent terms  will lead to $\log\delta_j$. There are actually  three $x$-independent terms in ${\cal N}^\lambda(q^2)$ -- those proportional to $y(1-y)$ -- localized in the form factors $a$, $e$ and $f$ which 
give rise the $\log\delta_j$ after the $\int \int \D x \D y\;[{\cal N}^\lambda(q^2) /{\cal D}_j(0)] $ integration. The constant $A$ in (8) comes from the form factor $a$ and the constant $B$ from the form factors $e, f$. The $x$-dependent terms in ${\cal N}^\lambda(q^2)$ yield  $\delta_j$, $\delta_j\log\delta_j$ and $\delta_j^2\log\delta_j$ terms, they are grouped into $G(m_j)$.   

The presence of this infrared divergence $\log\delta_j$ from diagrams similar to Fig.1 has been noticed in the literature$^{3, 4}$. We remark that the $q^2y(1-y)$ term (and not the $\log({\cal D}_j(q^2)/M_{\rm W}^2)$) in the form factor $a$  of (5) which  yields the important $\log\delta_j$.  The form factors $e $ and $f$ on the other hand, because of their $q^\lambda$ operator, contribute negligibly when contracted to the  lepton pair $\ell^+\ell ^-$  current. The dominant term of $\Gamma_j^\lambda(q^2)$ is found  to be

\begin{equation}
\Gamma^\lambda _j(q^2) ={\I g^3  \over 64 \pi^2 \cos\theta_{\W}} \left[ {q^2\over 6 M_{\rm W}^2}\right] \left( U_{\tau j}U_{\mu j}^*\log{m_j^2\over M_{\rm W}^2}\right)\gamma^\lambda (1-\gamma_5) \;.
\end{equation} 
From (9), the $\tau\to \mu +\ell^+ +\ell^-$ decay amplitude is 
\begin{equation}
{\cal A} (\tau\to \mu +\ell^+ +\ell^-) = {G_{\rm F}\over\sqrt{2}} 
\left({\alpha \over
24 \pi \sin^2\theta_{\W}}\right) {(M-m)^2\over M^2_{\rm W} -(M-m)^2 \cos^2
\theta_{\W} } 
\left( \sum_j U_{\tau j}U_{\mu j}^*\,\log \delta_j\right) \;L^\lambda \ell_\lambda
\end{equation} 
\begin{equation}
L^\lambda =\overline{u}(p)\gamma^\lambda(1-\gamma_5) u(P)\,, \; 
\ell_\lambda =\overline{u}(k_{-})\gamma_\lambda (g_V-\gamma_5 g_A) v(k_{+})\,, 
\; g_V={-1\over 2} +2 \sin^2\theta_{\W}\;,\; g_A = {-1\over 2} 
\end{equation} 
For ${\cal A} (\tau\to \mu +\rho^0)$, we simply replace $\ell_\lambda$ by $m_\rho 
f_\rho \varepsilon_\lambda$ where $f_\rho \approx 150$ MeV is the decay constant of the $\rho^0$, extracted from $\rho^0\to $ e$^+$ + e$^-$, and $\varepsilon_\lambda$ is the $\rho^0$ polarization vector. 
It remains to evaluate
\begin{equation}
{\cal B}=\sum_{j=1}^3 U_{\tau j} U_{\mu j}^* \log \left({m^2_j\over 
M_{\rm W}^2}\right) =
\sum_{
k=2}^3 U_{\tau k} U_{\mu k}^*  \log \left({m^2_k\over m^2_1}\right) 
\end{equation} 
where we have used $\sum_j U_{\mu j}^* U_{\tau j}=0$ to get rid of  the first $\nu_1$ mixing parameter
$U_{\mu 1}^* U_{\tau 1}$.
The factor ${\cal B} $ in (12) which represents the soft GIM cancelation
tells us that when  $m_j$ are degenerate, the lepton flavor mixing does not occur and the neutrinoless-$\tau$ decays identically vanish.

To estimate  ${\cal B}$, let us assume$^{(6)}$ the following form of the ${\cal U}_{\rm lep}$, neglecting possible CP violation in the lepton sector: 
\begin{equation}
{\cal U}_{\rm lep} = \pmatrix{ \cos\theta_{12} & -\sin\theta_{12}& 0 \cr 
{1\over \sqrt{2}}\sin\theta_{12} & {1\over \sqrt{2}}\cos\theta_{12}  & 
{-1\over \sqrt{2}}\cr
        {1\over \sqrt{2}}\sin\theta_{12}&{1\over \sqrt{2}}\cos\theta_{12}  & 
{1\over \sqrt{2}}\cr}\,.        
\label{eq:4} 
\end{equation}

The mixing angle $\theta_{23}\approx 45^0$ is suggested by the Super-Kamiokande data and
the $\theta_{13}\approx 0^0$ comes from the Chooz data$^{(6)}$ which give $\theta_{13} \leq 13^0$, whereas $\theta_{12}$ being arbitrary. Thus,
\begin{equation}
{\cal B}= \cos^2\theta_{12}\, \log {m_2\over m_3} +\sin^2 \theta_{12}\, \log {m_1\over m_3} \;.
\end{equation}
Although  $\theta_{12}$ is 
likely small $\approx 0^0$, however the maximal mixing $\theta_{12}\approx 45^0$ may be also possible allowing $\nu_{\rm e}\leftrightarrow \nu_\mu$ (as suggested by the LSND experiment).  
Taking $\theta_{12}$ in the range 
$0^0$--$45^0$, and using$^{(1,6)}$ $\Delta m^2_{23} =\vert m_3^2-m_2^2\vert =2\times 10^{-3}$ eV$^2$, $m_3 \approx 5\times 10^{-2}$ eV, then
 $\vert{\cal B}\vert^2$ is of order of unity 
[$\vert{\cal B}\vert^2 \sim {\cal O}(1)$], it could be bigger if $m_1$ or $m_2$ are exponentially smaller than $m_3$. We get the branching ratio  
\def \BB {\hbox {B}}
\begin{equation}
 {\BB}(\tau \to \mu +\ell^+ +\ell^-) \geq 10^{-14} \;.     
\label{eq:5} 
\end{equation}
Although (15) is still very small, however compared to the radiative case $\tau \to \mu +\gamma$, it represents a spectacular enhancement through the mild GIM cancelation.
 Finally, we remark that for large $q^2\geq M_{\rm W}^2$, for instance in Z$\to \mu^{\pm} + \tau^{\mp} $, we have the same ten diagrams with a real Z boson. However, obviously the $q^2 = M_{\rm Z}^2$ expansion does not make sense, and $F(M_{\rm Z}^2, m_j)$ in (8) cannot be approximated by $F(0, m_j) =
\log \delta_j$. Only for small $q^2\ll M_{\rm W}^2$ that the use of the dominant term $F(0, m_j)$ can be justified. Therefore the low energy e$^{+} +{\rm e}^{-} \to \mu^{\pm} + \tau^{\mp}$ reaction might be also worth to investigate. However compared to the one-photon exchange e$^{+} +{\rm e}^{-} \to \mu^{+} + \mu^{-}$ cross-section, the  ${\rm e}^{+} +{\rm e}^{-} \to \mu^{\pm} + \tau^{\mp}$ cross-section is damped, besides the coefficient ${\cal B}\alpha_{\rm em}/ 24 \pi \sin^2\theta_{\W}$ squared, by an additional $s^2/(M_{\rm Z}^2-s)^2$ multiplicative factor due to the Z propagator.

Independently of the precise numerical value of  ${\cal B}$, the neutrinoless decay modes  $\tau\to \mu +\rho^0$ and/or $ \tau\to \mu +\ell^+ +\ell^-$ are interesting on their own right for two reasons. First, although being higher order loop effect, the branching ratios are not desperately small due to the smoothly logarithmic GIM suppression. Second, while neutrino oscillations only provide the mass difference $\Delta m_{jk}^2$, lepton flavor changing $\tau$ decays give the ratio $m^2_j/m^2_k$. When combining these two processes, the absolute value of the neutrino mass $m_j$ may be obtained. Both reactions are mutually complementary in the determination of the neutrino masses and the lepton  mixing angles. 

Lepton flavor changing processes in the seesaw-type neutrino models are also discussed in$^{7}$. 

\hskip-0.6cm{\bf Acknowledgments :} 
Helpful discussions with John Iliopoulos are gratefully acknowledged. 
\newpage



\large
{\bf Figure Caption}  :  \normalsize 

Figures 1--3 : One-loop lepton flavor changing  $\tau \to \mu +\ell^+ + \ell^-$.


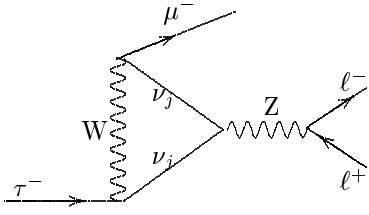
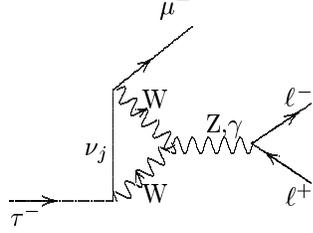
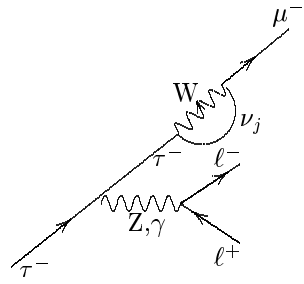
\end{document}